\documentstyle[11pt]{article}

\catcode`@=11
\@addtoreset{equation}{section}
\catcode`@=12

\def\vec#1{{\bf #1}}
\def\sub#1{_{{\rm #1}}}

\def\ind{_{\rm ind}}
\def\ext{_{\rm ext}}

\def\figgap#1#2#3{\begin{figure} \centerline{{} \hrulefill {}}
\begin{center} #1 \end{center}
\centerline{{}\hrulefill {\large\em Figure \ref{#3}} \hrulefill{}}
\caption[]{\small \sf #2}
\label{#3} \end{figure}}

\begin{document}

\title {\LARGE\bf The third electromagnetic  constant\\[2mm]
\LARGE\bf of an isotropic medium\bigskip}

\author{Jos\'e  F. Nieves\\
\normalsize \em Department of Physics,  University of
Puerto Rico, Rio Piedras, PR 00931, USA\\ \medskip \\ Palash B. Pal \\
\normalsize \em  Center for Particle Theory,
University of Texas, Austin, TX 78712, USA.}

\date{}
\maketitle

                \begin{abstract} \normalsize\noindent In addition to
the dielectric and magnetic permeability constants, another constant
is generally needed to describe the electrodynamic properties of a
linear isotropic medium.  We discuss why the need for the third
constant arises and what sort of physical situations can give rise to
a non-zero value for it. This additional constant, which we call the
{\em ``Activity Constant''} and denote by $\zeta$, can explain optical
activity and other phenomena from a purely macroscopic and
phenomenological point of view.
\end{abstract}\bigskip\bigskip

\section{Introduction}
                 In principle, the best way to discuss electromagnetic
processes taking place within a material medium is to use the
subatomic picture in which matter is but a huge number of fundamental
particles crowded in the vacuum. The charges and currents of these
particles can be fed into the equations of vacuum electrodynamics.
The resulting equations, if solved, can describe all electromagnetic
phenomena.

In practice, however, this is not a viable approach because, as
already commented, the number of charges in any macroscopic sample of
a material is enormous. One, therefore, has to take a simplified
approach. This is achieved by dividing all the sources of charge into
two parts: the external and the induced. The external charges are
described by constructing the charge- and current- densities of them
and feeding them into Maxwell's equations. The presence of the induced
charges is taken into account by introducing a few empirical constants
in the equations.

The number of such constants necessary depends on the nature of the
medium in question. For the simplest isotropic materials, it is well
known that one needs two quantities, the dielectric constant
$\epsilon$ and the magnetic permeability $\mu$, to describe all
phenomena of macroscopic electrodynamics \cite{text}.  However, for
certain isotropic systems one needs in general three constants (not
two) to describe electrodynamic phenomena in them.  While this fact is
also well known \cite{OpAct}, it is not widely discussed in standard
physics textbooks on electromagnetism \cite{physicsbooks}.

In this paper we explain, in simple terms, why the third constant
(which we denote by $\zeta$) arises and what kind of situations might
lead to the vanishing of this extra constant.  Some interesting
physics associated with $\zeta$ is discussed. This includes the
phenomena of optical activity of some active solutions, and of atomic
parity violation which arises due to the breaking of the parity
symmetry in the weak interactions.  Because $\zeta$ is crucial in
explaining these phenomena from a macroscopic point of view, we refer
to it as the {\em ``Activity constant''.} A discussion of the same
issues from a quantum field theoretic point of view can be found in
Ref.\ \cite{NP1} and Ref.~\cite{NP2}.

\section{Formulation of Macroscopic Electrodynamics} \label{formu}
              In the vacuum, Maxwell's equations consist of two
homogenous equations in the electric field $\vec E$ and the magnetic
field $\vec B$:
                \begin{eqnarray}
\nabla \cdot \vec B = 0  \quad , \quad
c \nabla \times \vec E = -\, {\partial \vec B \over \partial t}
\label{MaxEq1xt}
                \end{eqnarray}
and two inhomogeneous equations
involving the sources:
                \begin{eqnarray}
\nabla \cdot \vec E = \rho \quad , \quad
c \nabla \times \vec B - {\partial \vec E \over \partial t} =
\vec j \,. \label{MaxEq2xt}
                \end{eqnarray}
We have used the Heaviside-Lorentz
units here so that no factors of $4\pi$ appear in the equations. On
the other hand, Coulomb law and Biot-Savart law appear now with
factors of $(4\pi)^{-1}$. As usual, $c$ is a universal constant
denoting the magnitude of the velocity of light in the vacuum. For
rewriting any equation of the present paper in other popular systems
of units, see Table \ref{t:dimen}.

\begin{table} \caption[]
{{\small\sf Electromagnetic quantities and their dimensions, in
Heaviside-Lorentz (HL) units, in terms of the fundamental units of
length ($L$), time ($T$) and electric charge ($Q$) \cite{fn:dim}.  To
find the dimensions of the Fourier transforms of these quantities,
multiply any entry by $L^3 T$. In the last two columns, we give the
recipe to transform the quantities in other systems of
units.}\label{t:dimen}}

\begin{center}
\begin{tabular}{|c||l|c|c|} \hline
Quantity & Dimension &
\multicolumn{2} {c|} {Replace by the following to change to} \\
\cline{3-4} &in HL units & Gaussian units & SI units \\ \hline
$\rho$ & $QL^{-3}$ & $\sqrt{4\pi}\, \rho$ & $\rho/\sqrt{\epsilon_0}$
\\ $\vec j$ & $QL^{-2} T^{-1}$ & $\sqrt{4\pi}\, \vec j$ & $\vec j
/\sqrt{\epsilon_0}$\\ $\vec E$ & $QL^{-2}$ & $\vec E /\sqrt{4\pi}$ &
$\sqrt{\epsilon_0} \,
\vec E$\\
$\vec B$ & $QL^{-2}$ & $\vec B /\sqrt{4\pi}$ & $\vec B /\sqrt{\mu_0}$
\\ $c$ & $LT^{-1}$ & $c$ & $1/\sqrt{\mu_0 \epsilon_0}$ \\ \hline
\end{tabular}
\end{center}
\end{table}
                      For future purposes, let us note the dimensions
of various electromagnetic quantities in these units. Of course,
$\rho$ has the dimension of $QL^{-3}$ since, by definition, it is the
charge per unit volume \cite{fn:dim}. Once that is known, Eqs.\
(\ref{MaxEq1xt}) and (\ref{MaxEq2xt}) determine the dimensions of
$\vec E$, $\vec B$ and $\vec j$. The results are summarized in Table
\ref{t:dimen}.

It is convenient to re-write all the above equations in terms of their
Fourier components. We therefore express every quantity $\Phi(\vec
x,t)$ as a superposition of plane waves through the relation
\cite{fn:fourier}
                \begin{eqnarray}
\Phi(\vec x,t) = {1 \over (2\pi)^2} \int d\omega \int d^3 \vec k \,
\widetilde{\Phi} (\vec k, \omega) \exp [i(\vec k \cdot \vec x -
\omega t)] \,.   \label{fourier}
                    \end{eqnarray}
{}From the usual results of Fourier
analysis, it is then easy to show that
                \begin{eqnarray}
\nabla \rightarrow i\vec k , \quad \partial/\partial t \rightarrow
-i\omega \,, \label{derivFouri}
                \end{eqnarray}
which indicates how the
derivatives on co-ordinate space functions affect the Fourier
components. With the aid of these properties, we can cast Maxwell
equations in the form
                \begin{eqnarray}
\vec k \cdot \vec B =0 \quad&,&\quad c \vec k \times \vec E =
\omega \vec B  \label{MaxEq1kw} \\
i\vec k \cdot \vec E = \rho \quad &,& \quad ic \vec k \times
\vec B + i\omega  \vec E = \vec j \,.
\label{MaxEq2kw}
                    \end{eqnarray}
This form makes the ensuing
discussion much easier since the equations become algebraic. This is
especially convenient later when we introduce the dielectric constant
$\epsilon$ and the magnetic permeability $\mu$ of a medium through
Eqs. (\ref{P&M}) and (\ref{epsmu}), which would become integral
relations in co-ordinate space. Notice that we write the Fourier
components with the same notation as the space-time functions
themselves. This will not cause any confusion because the space-time
functions will never again be used in this article.

As mentioned in the Introduction, the discussion of macroscopic
electrodynamics starts from the observation that $\rho$ and $\vec j$
contain, in addition to any external sources (denoted by the subscript
``ext''), the sources induced in the medium (denoted by the subscript
``ind''):
                \begin{eqnarray}
\rho = \rho\ext  + \rho\ind \quad, \quad
\vec j = \vec j\ext + \vec j\ind \,. \label{ext+ind}
                    \end{eqnarray}
Ordinarily, the induced sources are
expressed in terms of the polarization ($\vec P$) and the
magnetization ($\vec M$) as follows:
                \begin{eqnarray}
\rho \ind &=& -i\vec k \cdot \vec P \nonumber\\
\vec j\ind &=& -i\omega \vec P + ic \vec k \times \vec M \,.
\label{indparam}
                    \end{eqnarray}
In this form, the expressions for
$\rho \ind$ and $\vec j\ind$ mimic those for total $\rho$ and $\vec j$
given in Eq.\ (\ref{MaxEq2kw}), and they automatically satisfy the
equation of continuity, which takes the following form in Fourier
space:
                \begin{eqnarray}
\vec k \cdot \vec j \sub{x} - \omega \rho \sub{x} =0 \,,
\label{currconsv}
                    \end{eqnarray}
where the subscript `x' stands for
the induced sources. This signifies the conservation of induced
charges. The conservation equation (\ref{currconsv}) also holds when
`x' stands for the total or just the external charges.

The dielectric constant $\epsilon$ and the magnetic permeability $\mu$
for a system are now introduced by writing
                \begin{eqnarray}
\vec P = (\epsilon -1) \vec E \quad,\quad \vec M = (1 - \mu^{-1})
\vec B \,.
\label{P&M}
                    \end{eqnarray}
In writing these relations it is
assumed, as we do in this paper, that the response of the system to
applied electromagnetic fields is linear.  This excludes special
systems such as ferroelectrics and ferromagnetics from consideration,
but otherwise does not impose any real restriction provided that the
applied fields are not extremely large.  Further, we will consider the
simple case of isotropic media, for which $\epsilon$ and $\mu$ can be
taken as scalars \cite{fn:tensor}. One can then use Eqs.
(\ref{ext+ind}), (\ref{indparam}) and (\ref{P&M}) to rewrite Eqs.\ (\ref
{MaxEq1kw}) and (\ref{MaxEq2kw}) as
                \begin{eqnarray}
\vec k \cdot \vec B =0 \quad &,&\quad c \vec k \times \vec E =
\omega \vec B\label{MaxEq:hom}  \\
i\vec k \cdot \vec D = \rho\ext \quad&,&\quad ic \vec k \times \vec H
+ i\omega \vec D = \vec j\ext \,.
\label{MaxEq:DH}
                    \end{eqnarray}
where
                \begin{eqnarray}
\vec D \equiv \vec E + \vec P = \epsilon \vec E \quad,\quad
\vec H = \vec B - \vec M = \mu^{-1} \vec B \,.
\label{epsmu}
                    \end{eqnarray}

Since the source terms are just the external charges and
currents, one can in principle solve Eq.\ (\ref {MaxEq:DH}) for $\vec
D$ and $\vec H$.  Once the empirical constants $\epsilon$ and $\mu$
are known, one can then use Eq.\ (\ref {epsmu}) to derive $\vec E$ and
$\vec B$.

In the next section, we will see that this treatment makes an
important assumption tacitly, and because of that, it cannot describe
an important class of physical phenomena within the framework of
macroscopic electrodynamics.

\section{The most general parametrization of medium properties}
\subsection{Parametrization of the induced sources}
              The crucial question in macroscopic electrodynamics is:
how to parametrize the induced sources $\rho\ind$ and $\vec j\ind$. It
is important to recognize that these quantities cannot be determined
from considerations of macroscopic physics alone.  In standard
textbooks of electromagnetism, some simple microscopic models of the
medium are invoked to express $\vec P$ and $\vec M$ in terms of
microscopic quantities, and thereby motivate the parametrization
discussed above.  However, such models may not describe all possible
media and, therefore, the expressions in Eq.  (\ref{indparam}) need
not be the most general ones. From a purely macroscopic point of view,
we can try to parametrize $\rho\ind$ and $\vec j\ind$ in the most
general form subject to the conditions of isotropy and linearity. As
we show below, this approach indicates that Eq. (\ref{indparam})
indeed does not provide the most general expressions for the induced
charge and current densities.

To clarify the point, we summarize Eqs. (\ref{indparam}) and (\ref{P&M})
by writing
                \begin{eqnarray}
\rho \ind &=& -\,i (\epsilon-1) \vec k \cdot \vec E   \label{rhoind} \\
\vec j \ind &=& -i (\epsilon -1) \omega \vec E + ic (1 - \mu^{-1})
\vec k \times \vec B \,. \label{jind2}
                    \end{eqnarray}
Looked at this way, it seems that
we are trying to write down $\rho\ind$ and $\vec j\ind$ as linear
functions of the electric and the magnetic fields, where the
coefficients depend on $\omega$ and $\vec k$.  One can then ask, are
the combinations the most general possible ones subject to the
assumption of isotropy?

The answer is `yes' for the expression for $\rho \ind$ in Eq.\ (\ref
{rhoind}).  Note that $\rho$ is a scalar as far as rotations are
concerned, and the only scalars which are linear functions of $\vec E$
and $\vec B$ are $\vec k \cdot \vec E$ and $\vec k
\cdot \vec B$. Of them, the latter  is identically zero (Eq.\
(\ref{MaxEq1kw})), so that $\rho \ind$ has to be proportional to $\vec
k \cdot \vec E$, as shown in Eq.\ (\ref {rhoind}).

The situation is different for Eq.\ (\ref {jind2}). Here, $\vec j
\ind$ is a vector,  and there are four vectors involving $\vec E$ and
$\vec B$, viz., $\omega\vec E$, $c\vec k \times \vec B$, $\omega\vec
B$ and $c\vec k \times \vec E$, which have the same dimensions as
$\vec j$. The last two are equal by Eq.\ (\ref {MaxEq1kw}), so that
only the first three can be taken as independent. One can therefore
write the most general expression for $\vec j$ as
                \begin{eqnarray}
\vec j \ind &=&
-i (\epsilon -1) \omega \vec E + ic (1 - \mu^{-1}) \vec k \times
\vec B -i \zeta \omega \vec B\,. \label{jind-general}
                    \end{eqnarray}
Notice that the coefficient in the
first term is fixed to be the same as the coefficient in Eq.\
(\ref{rhoind}) by current conservation. Thus, in general, we have
three dimensionless constants describing the electromagnetic
properties of an isotropic medium: $\epsilon$, $\mu$ and $\zeta$.

Looking back at Eq.\ (\ref {jind2}), it now becomes immediately
obvious why it is incomplete. Since $\vec j\ind$ is a vector having
three components, one needs, in general, three basis vectors to write
it down. In Eq.\ (\ref {jind2}), we used only two basis vectors,
$\omega\vec E$ and $\vec k \times \vec B$. Therefore, Eq.\
(\ref{jind2}) is valid only under some extra assumptions, the precise
nature of which will be discussed in the next section.

One comment is in order. The last equation in Eq.\ (\ref {MaxEq2kw})
expresses the total current $\vec j$ as a combination of the two
vectors $\omega\vec E$ and $\vec k \times \vec B$. If we have had no
complaint about that, why do we argue about Eq.\ (\ref{jind2}), which
expresses $\vec j\ind$ as a linear combination of the same two
vectors?

The answer to this question lies in the difference of the nature of
Eqs. (\ref{MaxEq2kw}) and (\ref{jind2}).  Eq.\ (\ref {MaxEq2kw}) is a
{\em law}, describing the fundamental nature of electromagnetic
interactions.  Eq.\ (\ref {jind2}) is a {\em parametrization} of the
induced current, which expresses our ignorance about the details of
the dynamics of all processes going on in the medium or our inability
to quantify them in minutest details.  Such parametrization must be as
general as possible subject to the conditions of the problem, and Eq.\
(\ref {jind-general}) provides this general form for $\vec j\ind$.

With this parametrization of the induced sources, the pair of
inhomogeneous equations given in Eq.\ (\ref{MaxEq2kw}) become
                \begin{eqnarray}
\label{Max:inhom}
i\epsilon\vec k\cdot\vec E & = & \rho\ext \,,\nonumber\\
i\epsilon\omega\vec E + i\frac{c}{\mu}\vec k\times\vec B +
i\zeta\omega \vec B & = & \vec j\ext \,,
                \end{eqnarray}
while the
sourceless equations (\ref{MaxEq1kw}) remain intact.  These two
equations plus the equations in (\ref{MaxEq1kw}) can be taken as the
fundamental equations for the $\vec E$ and $\vec B$ fields in a medium
in the presence of external sources.  In the next sections we will
elucidate some of the physical implications and consequences
associated with the constant $\zeta$.

\subsection{Constitutive relations}
        As already mentioned in the Introduction, the inadequacy of
the simple constitutive relations given in Eq.\ (\ref{epsmu}) for
describing a certain class of physical phenomena has been recognized
for a long time \cite{OpAct}.  In many instances this deficiency has
been addressed \cite{OpAct,Monzon,Krowne,Kong} by postulating Eqs.\
(\ref{MaxEq:hom}) and (\ref{MaxEq:DH}) together with a more general
set of constitutive relations to replace those in
Eq.\ (\ref {epsmu}):
                \begin{eqnarray}
\vec D &=& \epsilon \vec E + \beta \vec B  \,, \label{D=EB}\\
\vec H &=& \gamma \vec E + \mu^{-1} \vec B \,. \label{H=EB}
                \end{eqnarray}
With this approach, it seems that one needs two
extra constants for the medium, $\beta$ and $\gamma$, rather than the
constant $\zeta$ discussed earlier.  However, it is easy to see that
the two constants introduced here are not independent. For this,
substitute Eqs.\ (\ref{D=EB}) and (\ref{H=EB}) into Eq.\
(\ref{MaxEq:DH}). This yields
                \begin{eqnarray} i\epsilon\vec
k\cdot\vec E & = & \rho\ext\,,\nonumber\\ i\frac{c}{\mu}\vec
k\times\vec B + i\epsilon \omega \vec E + i(\beta + \gamma) \omega
\vec B & = & \vec j\ext\,,
                \end{eqnarray}
Thus, it is only the sum
$\beta + \gamma$ that appears in the field equations, and therefore
only three constants have an independent physical significance.
Moreover, comparing these equations with those in Eq.\
(\ref{Max:inhom}), we see that $\beta + \gamma = \zeta$.  Because of
this redundancy, one can try to put an arbitrary constraint to fix $\beta$ and
$\gamma$, e.g., one might put \cite{Tellegen} $\beta=\gamma$.

While the approach of postulating more general constitutive relations
rather than the simple ones given in Eq.\ (\ref{epsmu}) has been
followed in the literature for a long time, the lesson we have learned
from the above discussion is that, in general, the constitutive
relations that have been used contain redundant parameters.
The reason for the redundancy is easy to understand. The
componets of $\vec E$ and $\vec B$ are not independent. They are
constrained by Eq.\ (\ref{MaxEq1kw}).
Similarly, the components of $\vec D$ and $\vec H$ also are not
independent. The constitutive relations, being relations between these
two sets of variables none of which are independent, can also have
redundant parameters. This can also be seen from Eq.\
(\ref{MaxEq:DH}). In fact, only the equation for $\vec j\ext$ is
independent there; the equation for $\rho\ext$ can be derived from it
using the charge conservation equation, Eq.\ (\ref{currconsv}).
Taking the general expression for $\vec j\ext$ from Eq.\
(\ref{Max:inhom}) and eliminating $\vec B$ using Faraday's law, Eq.\
(\ref{MaxEq:hom}), the equation for $j\ext$ in Eq.\ (\ref{MaxEq:DH})
becomes
                \begin{eqnarray} c \vec k
\times \vec H + \omega \vec D =
\epsilon\omega\vec E + \frac{c^2}{\omega\mu}\vec k\times (\vec k
\times \vec E) + c\zeta \vec k \times \vec E
                \end{eqnarray}
Thus, we are trying to define two vectors $\vec
H$ and $\vec D$ from this single equation, in terms of just $\vec E$,
which is the only independent variable here. There is no unique way to
do this, and any parametrization will contain some arbitrariness.  In
contrast, the components of $\vec j\ind$ are all independent. By
focusing the attention on them, we have succeeded in finding a
parametrization of the induced sources that is both minimal and
complete.  Because of this redundancy of the
parametrization in terms of the constitutive relations, it is more
convenient to use the parametrization of $\vec j\ind$ instead, which
we will do in the rest of this article.

We should comment that the restriction of the number of
electromagnetic constants to three applies only in the Fourier space,
which we use in the present article. In co-ordinate space on the other
hand, one may add many terms using higher and higher derivatives of
the $\vec E$ and $\vec B$ fields in the definition of the induced
current \cite{Satten}. Translated to the Fourier space, those
additional terms are taken into account by allowing the constants
$\epsilon$, $\mu$ and $\zeta$ to depend on higher powers of $\vec k$
and $\omega$.

\section{The constant $\zeta$ and the discrete symmetries of
space-time}                            Why is it that the constant
$\zeta$ is not introduced for
the simplest isotropic materials?  Alternatively, what are the
conditions under which $\zeta$ can have a nonzero value? The key to
the answer to these questions lies in the discrete symmetries of space
and time, as we discuss in this section.

\subsection{Definition of the discrete symmetries P, T, C}
                 For orientation, consider Newton's law for a particle
of mass $m$ moving under the influence of the gravitational force
exerted by another particle of mass $M$ placed at the origin.  The
equation of motion is
                \begin{eqnarray}
\label{eqmotion}
m \vec a = -\, {GMm \over r^3} \, \vec r,
                \end{eqnarray}
where the
right hand side denotes the gravitational force on the particle.  If
we perform space inversion (i.e., change $\vec r$ to $-\vec r$), the
force changes sign, but so does $\bf a$, since it is defined to be
$d^2 \vec r/dt^2$. So Eq.\ (\ref {eqmotion}) does not change.  Similarly, if
we perform a time inversion (i.e., change $t$ to $-t$), neither side
of Eq.\ (\ref {eqmotion}) changes so, again, the equation remains invariant.

\begin{table} \caption[]
{{\small\sf The transformation properties of various physical
quantities under the operations P, T and C. The quantity $\eta_P$, for
example, is defined in Eq. (\ref{PhiP}), and the definitions of
$\eta_T$, $\eta_C$ etc are similar. The last two columns also show the
transformation under the combined operations CP and CPT.}\label{t:CPT}}
\begin{center} \begin{tabular}{|c||c|c|c|c|c|} \hline &
\multicolumn{5}{c|}{Transformation given by} \\ \hline Quantity &
$\eta_P$ & $\eta_T$ & $\eta_C$ & $\eta_{CP}$ & $\eta_{CPT}$ \\ \hline
$\vec E$ & $-$ & $+$ & $-$ & $+$ & $+$ \\ $\vec B$ & $+$ & $-$ & $-$ &
$-$ & $+$ \\ $\vec k$ & $-$ & $+$ & $+$ & $-$ & $-$ \\ $\omega$ & $+$
& $-$ & $+$ & $+$ & $-$ \\ $\rho$ & $+$ & $+$ & $-$ & $-$ & $-$ \\
$\vec j$ & $-$ & $-$ & $-$ & $+$ & $-$ \\ $\vec A$ & $-$ & $-$ & $-$ &
$+$ & $-$ \\
\hline
  \end{tabular} \end{center}
\end{table}

For a long time, it was believed that these two discrete operations,
viz. of space inversion or parity (P) and time reversal (T), do not
change any laws of physics. If this were true, all the terms in a
physically acceptable equation would transform the same way under P
and T, as in the example discussed above.  Indeed, we can verify that
such is the case with the Maxwell equations in (\ref{MaxEq1xt}) and
(\ref{MaxEq2xt}) or, equivalently, with their Fourier space
counterparts given in Eqs.\ (\ref{MaxEq1kw}) and (\ref{MaxEq2kw}).  To
show this, we proceed as follows.

Consider, for example, Parity.  The identification given in Eq.\
(\ref{derivFouri}), implies that in Fourier space the parity operation
is induced by the substitution
                \begin{eqnarray}
(\vec k,\omega) \stackrel{P}{\longrightarrow} (-\vec k,\omega)
                \end{eqnarray}
It is very easy to verify that, if we simultanously make the following
substitutions
                \begin{eqnarray}
\label{PclassEBj}
\vec E (\vec k, \omega) &\stackrel{P}{\longrightarrow} &
- \vec E (\vec k, \omega) \,,
\nonumber\\
\vec B (\vec k, \omega) &\stackrel{P}{\longrightarrow} &
\phantom{-} \vec B (\vec k, \omega) \,,
\nonumber\\
\vec j (\vec k, \omega) & \stackrel{P}{\longrightarrow} &
-\vec j (\vec k, \omega) \,,
\nonumber\\
\rho (\vec k, \omega) & \stackrel{P}{\longrightarrow} &
\phantom{-} \rho (\vec k, \omega) \,,
                                \end{eqnarray}
then the equations are
not changed.  If for every quantity, including $\omega$ and $\vec k$
themselves, we write the above transformation rules in the form
                \begin{equation}
\label{PhiP}
\Phi \left( \vec k, \, \omega \right)
\stackrel{P}{\longrightarrow} \eta_P \Phi \left(\vec k, \,\omega
\right)\, ,
                                \end{equation}
then the multiplicative
constant $\eta_P$ has the value given in Table \ref{t:CPT} for the
various quantities.

The same procedure is followed for the Time reversal symmetry, for
which Eq.\ (\ref{derivFouri}) suggests \cite{fn:T}
                \begin{eqnarray}
(\vec k,\omega) \stackrel{T}{\longrightarrow} (\vec k,-\omega) \,.
                \end{eqnarray}
The fact that the Maxwell equations are invariant under
P and T further bolstered the idea that all the laws of physics are
invariant under those operations.

It is easily seen that the Maxwell equations (Eqs. (\ref{MaxEq1xt})
and (\ref{MaxEq2xt}), and their Fourier space version given in Eqs.\
(\ref{MaxEq1kw}) and (\ref{MaxEq2kw})) have another discrete symmetry
since they are unaffected by the operation under which $\rho$, $\vec
j$, $\vec E$ and $\vec B$ all change sign.  This operation, called
charge conjugation (C) because the sign of all the charges are
reversed, plays an important role in quantum field theory, where it
converts a particle into its associated antiparticle.

\subsection{Implications of P, T, C symmetries on $\epsilon$,
$\mu$, $\zeta$ } Now that we know the transformation properties of the
electromagnetic fields and currents, we can look back at Eq.\
(\ref{jind-general}) with respect to the discrete symmetries C, P and
T, and some important combinations like CP and CPT.  To proceed we
first notice that we can consider instead
                \begin{eqnarray}
\label{jext}
\vec j\ext (\vec k,\omega) =
i \epsilon(k,\omega) \omega \vec E(\vec k,\omega) +
\frac{c}{\mu(k,\omega)} i \vec k \times \vec B(\vec k,\omega)
 + i \zeta(k,\omega) \omega \vec B(\vec k,\omega),
                \end{eqnarray}
which is equivalent to Eq.\ (\ref{jind-general}) because of the
definition in Eq.\ (\ref {ext+ind}). We have explicitly indicated the
dependence on $\vec k$ and $\omega$ of all the quantities involved,
because it is important to keep that in mind in what follows. In
addition, we have explicitly used the fact that the quantities
$\epsilon$, $\mu$ and $\zeta$, which represent properties of the
medium, can depend on $\vec k$ only through its magnitude $k
\equiv\left|\vec k\right|$ since we are considering an isotropic
medium.

Before we proceed any further, we want to note that there is one
relation that the electromagnetic constants $\epsilon$, $\mu$ and
$\zeta$ must satisfy based on very general grounds.  It is simply the
requirement that the fields $\vec E$, $\vec B$ and the current density
$\vec j$ are, in coordinate space, real quantities. This implies that,
for real values of $\omega$, the electromagnetic constants satisfy
                \begin{eqnarray}
\label{hermiticity1}
f^\ast(-\vec k,-\omega) = f(\vec k, \omega) \,,
                \end{eqnarray}
where $f$ stands for any of them. As noted earlier, the dependence on
$\vec k$ can be only through its magnitude $k$ since we are
restricting the discussion to isotropic materials. Thus, in our case,
we can write
                \begin{eqnarray}
\label{hermiticity}
f^\ast(k,-\omega) = f(k, \omega) \,,
                \end{eqnarray}
which means that
the real parts of these electromagnetic constants are even in $\omega$
whereas the imaginary parts are odd.  We stress that this is a very
general property, not tied up in any way with the discrete symmetry
operations that we are going to discuss now. The discrete symmetries,
if they are valid in any given situation, impose extra conditions on
the electromagnetic constants.

Consider, for example, Parity.  The question we want to answer is the
following: If we take each quantity that appears in Table~\ref{t:CPT},
including $\vec k$ and $\omega$, and make in Eq.\ (\ref{jext}) the
replacement indicated in Eq.\ (\ref{PhiP}), what relations must be
satisified by $\epsilon$, $\mu$ and $\zeta$ in order for that equation
to remain the same?  In this fashion we easily derive the consequences
of the parity symmetry on $\epsilon$, $\mu$ and $\zeta$, and similarly
with the other discrete symmetries.  We discuss them one by one.

\paragraph*{Parity~:}
                 The parity reversed form of Eq.\ (\ref{jext}) is
                \begin{eqnarray}
- \vec j\ext = - i \epsilon \omega \vec E - {c
\over\mu} i\vec k \times \vec B + i \zeta \omega \vec B
                \end{eqnarray}
or
                \begin{eqnarray}
\vec j\ext  =  i \epsilon \omega \vec E + {c \over\mu} i\vec k
\times \vec B - i \zeta \omega \vec B \,. \label{jext/P}
                    \end{eqnarray}
Here and below, the arguments of
$\epsilon$, $\mu$, and $\zeta$, unless otherwise given, should be
assumed to be $k$ and $\omega$.  Comparing Eqs.\ (\ref{jext}) and
(\ref{jext/P}), we conclude that
                \begin{eqnarray}
\mbox{Parity symmetry} \Rightarrow \zeta = 0 \,.
                    \end{eqnarray}

\paragraph*{Charge Conjugation~:}
            The charge conjugated form of Eq. (\ref{jext}) is the same
as the original one since each terms changes sign, and therefore we do
not get any constraints on $\epsilon$, $\mu$, and $\zeta$.

\paragraph*{CP~:}
            Obviously, this gives the same constraint as parity alone
does, since charge conjugation does not affect Eq.\ (\ref {jext}), as
stated above.

\paragraph*{Time reversal~:}
The time-reversed form of Eq.\ (\ref{jext}) is
                \begin{eqnarray} - \vec
j\ext = - i \epsilon \omega \left( k,-\omega
\right) \vec E - {c \over \mu \left( k,-\omega
\right)} i\vec k \times \vec B +  i \zeta \omega \left( k,
-\omega \right) \vec B, \label{jext/t}
                \end{eqnarray}
which gives, on comparison with Eq.\ (\ref {jext}), the following constraints:
                \begin{eqnarray}
\mbox{Time reversal symmetry} \Rightarrow  \left\{ \begin{array}{rcl}
\epsilon \left( k,-\omega \right) &=& \epsilon \left( k,\omega
\right)\\
\mu \left( k,-\omega \right) &=& \mu \left( k,\omega
\right)\\
\zeta \left( k,-\omega \right) &=& - \zeta \left( k,\omega
\right)
\end{array} \right. \quad .
\label{Trules}
           \end{eqnarray}
Taken together with Eq.\
(\ref{hermiticity}), this implies that $\zeta$ must be purely
imaginary and an odd function of $\omega$, whereas $\epsilon$ and
$\mu$ should be real and even functions of $\omega$.

\paragraph*{CPT~:}
In this case, we get similarly the following constraints
                \begin{eqnarray}
\mbox{CPT symmetry} \Rightarrow
\left\{   \begin{array}{rcl}
\epsilon \left( k,-\omega \right)&=&\epsilon \left(
k,\omega \right)\\
\mu \left( k,-\omega \right)&=&\mu \left( k,\omega
\right)\\
\zeta \left( k,-\omega \right)&=& \zeta \left( k,\omega
\right)
\end{array}  \right. \quad .
           \end{eqnarray}

\subsection{Constraints on $\epsilon$, $\mu$, $\zeta$ in ordinary
media} From the relations deduced above, it is quite clear that the
presence of the $\zeta$-term implies some properties that are
asymmetric under P and CP transformations.  Similarly, since T and CPT
imply mutually inconsistent constraints on $\zeta$, at least one of
them must be violated in order for $\zeta$ to exist.

We said earlier that classically it was believed that all interactions
conserve P, T and C.  From this, one might have expected $\zeta$ to
vanish.  However, we know now that P, T and C are all violated in
fundamental interactions between particles.

Since 1957, we know that parity is violated by weak interactions.
Since 1964, we know that CP is also violated by the same, although to
a much lesser extent. Since $\vec j\ind$ arises because of complicated
processes taking place within the medium, including weak interaction
processes, there is no reason why the $\zeta$-term should not be
present in Eq.\ (\ref {jind-general}).

If we leave the discussion at that point, it might seem that the
parameter $\zeta$ can at best be very small, since the weak
interaction is indeed very feeble. What could be worse, the
$\zeta$-term violates CP, and the CP-violating effects are known to be
smaller than the usual weak interaction strength by about three orders
of magnitude. So $\zeta$, if present, would seem to be extremely tiny.
But that need not be true in general, as we argue shortly.

If we had carried the previous line of argument further, we would have
concluded that the only constraint on the value of $\zeta$ is $\zeta
\left( k,\omega \right) = \zeta \left( k, -\omega \right)$ which
follows from CPT symmetry, since there are very strong reasons to
believe that fundamental interactions between particles can never
violate CPT.  This again, would have been wrong for the reason
discussed below.

In principle, the quantities $\epsilon$, $\mu$, $\zeta$ could be
determined if we could solve the equations of motion of the particles
that compose the medium. These equations include the effects of the
mutual interactions between the particles as well as the effects of
the local $\vec E$ and $\vec B$ fields. These are not external
prescribed fields but are the fields that enter into Maxwell's
equations, which must also be solved for, and which in turn contain
the induced charge and current densities in the source terms. Thus,
Maxwell's equations plus the equations of motion of the particles in
the medium form a closed system which must be solved
self-consistently. To date, no one has succeeded in solving this
formidable problem and in practice one turns to a more modest
approach.

In classical as well as quantum physics, the calculation of
$\epsilon$, $\mu$, $\zeta$ proceeds in two steps. In the first, the
equation of motion of a single particle is set up, which includes the
effects of the local fields, as well as the effects of the other
particles in the medium.  From this, the contribution of a single
particle to the induced charge and current is obtained. In the second
step, this single-particle contribution is suitably averaged over the
particles in the medium, thereby obtaining an expression for the
induced charge and current densities, from which the values for
$\epsilon$, $\mu$, $\zeta$ can be extracted.

It is now clear that the asymmetries with respect to any operation can
enter into any quantity in either of these two steps. In the first
step, it enters if the underlying interactions do not respect the
symmetry generated by the operation, as is the case of weak
interactions with respect to the symmetries of P, C and CP. But the
asymmetries can also creep into the process of taking the average over
the particles in the medium, because the medium might have a surplus
of one kind of particles over others.

As a well-known example, consider the mass of the electron and its
antiparticle, the positron. In vacuum, the two masses must be equal.
Formally, this is ensured by the combined symmetry CPT.  Inside a
chunk of normal matter, however, there are many many electrons and
hardly any positrons, so that the interactions of the two particles
become very different. As a result, they move with different effective
masses within a medium. Thus, the CPT breaking effects creep in
through the asymmetry in the medium.

In the case of our present interest, considerations of CPT and CP
symmetries are irrelevant for the same reason.  In fact, the CP and
CPT asymmetries are large since normal matter does not contain any
antiparticles at all, and hence neither CP nor CPT stand in the way of
obtaining a large $\zeta$.

Let us now remind ourselves that in order to obtain a large $\zeta$,
we also need a large P-asymmetry.  The P-symmetry is of course
violated by weak interactions, and that can give rise to a small value
of $\zeta$. To obtain a large $\zeta$, we need a large P-asymmetry in
the medium.

Take a free electron gas, for example \cite{fn:proton}.
An electron in this gas can be specified by its momentum $\bf p$ and
spin angular momentum $\bf s$.  Under the action of parity, $\vec p
\rightarrow - \vec p$, $\vec s \rightarrow \vec s$.  However, in the
metal, there are as many electrons going in the $+x$ direction, say,
as in the $-x$ direction .  So the parity operation does not affect
the overall momentum distribution of the metallic electrons.  In other
words, this medium does not have any P-asymmetry and so $\zeta$ must
arise in this medium through parity violation in weak interaction.

Similarly, under the T operation, $
\vec p \rightarrow - \vec p, \quad \vec s \rightarrow  - \vec s.
$ But again, since the distributions of both $\bf p$ and $\bf s$ were
random to start with, the overall distributions do not change by this
operation, i.e., the medium is T-symmetric.  Therefore, barring the
small T-violating effects which are $10^{-3}$ times weaker than the
P-violating effects in weak interactions, the consequences of
T-symmetry should be taken seriously.  As we found before in Eq.\
(\ref{Trules}), this implies that $\zeta$ must be purely imaginary and
an odd function of $\omega$.

This is quite striking.  Note that $\epsilon$ and $\mu$ are both even
functions of $\omega$, so that it is conceivable to think of a medium
where they are constants.  In this case the medium is non-dispersive.
But $\zeta$ cannot have a nonzero constant term and therefore all
media are dispersive when it comes to $\zeta$.  Of course, we
emphasize that this comment is true only to the extent that
T-violation in fundamental interactions can be neglected.

When we consider more complicated media, it is possible to obtain
P-asymmetry from the medium itself.  A well-known example is that of
helical molecules, with one helicity in excess over the other. This
occurs in natural sugar solutions, for example, where in fact one
helicity is absent altogether.  If we consider the passage of
electromagnetic waves through such a solution in which the helical
axes are randomly oriented, and if the wavelength is larger than the
molecular dimension so that the wave cannot ``see'' individual
molecules, we have an isotropic medium with a built-in P-asymmetry.
This gives rise to large $\zeta$-term, and physically it manifests
itself as the phenomenon of optical activity, as we discuss in
Sec.~\ref{implications}.

In short, there is no reason why $\zeta$ should be zero in normal
matter, and its value can be large if there exist large asymmetries in
the medium. Therefore it is imperative that we understand its physical
significance.

\section{Dispersion relations}
         It is well-known that a great deal about the nature of
wave-propagation is learned from the dispersion relations. These are
the relations that $\omega$ and $\vec k$ must satisfy in order
that solutions to Eq.\ (\ref{Max:inhom})
exist when $\rho \ext = 0$ and $\vec j \ext = 0$; i.e.,
                \begin{eqnarray}
& \epsilon \vec k \cdot \vec E = 0 \label{rhoext=0} \\
& \epsilon \omega \vec E +
{\textstyle c\over \textstyle\mu} \vec k
\times \vec B +  \zeta \omega \vec B = 0 \,.
\label{jext=0}
                    \end{eqnarray}
These two, plus the two equations
in Eq.\ (\ref {MaxEq1kw}), can be taken as the fundamental equations
governing the $\vec E$ and $\vec B$ fields in a medium in the absence
of external charges and currents.

\figgap {
\begin{picture}(60,60)(25,25)
\thicklines
\put(50,50){\vector(1,0){30}}
\put(50,50){\vector(0,1){30}}
\put(50,50){\vector(-1,-1){20}}
\put(70,45){\makebox(0,0){{\large$\widehat{e}_2$}}}
\put(42,35){\makebox(0,0){{\large$\widehat{e}_1$}}}
\put(53,70){\makebox(0,0){{\large$\widehat{k}$}}}
\end{picture}}
{The relative orientations of the two transverse polarization vectors
with respect to the wave vector.} {orien} In the vacuum, since
$\epsilon = \mu = 1$ and $\zeta = 0$, Eq.\ (\ref {rhoext=0}) clearly shows
that $\vec E$ must be perpendicular to the direction of propagation of
the electromagnetic wave. Thus, one can discuss electromagnetic waves
in terms of two basis polarization vectors, $\hat e_1$ and $\hat e_2$,
both being perpendicular to $\vec k$, as shown in Fig.~\ref{orien}.

In the medium, Eqs.\ (\ref{rhoext=0}) and (\ref{jext=0}) can be satisfied for
a longitudinal electric field if $\vec B = 0$ and $\omega$ is such
that $\epsilon = 0$.  Since the physical properties of this
longitudinal mode depend on the expression for $\epsilon$ only, we
focus our attention on the transverse modes in order to discuss the
physical implications of $\zeta$.  Thus, eliminating $\vec B$ from
Eq.\ (\ref{jext=0}) by the use of Faraday's law and using $\vec
k\cdot\vec E = 0$, we obtain
                \begin{eqnarray}
\omega^2 \vec E = {c^2 \over \epsilon\mu} k^2 \vec E -
{c\zeta \over \epsilon} \omega k \hat k \times \vec E \,,
\label{waveeq}
                    \end{eqnarray}
where $\hat k = \vec k/k$. Let us
first see what happens if $\zeta = 0$, which is the case treated in
standard textbooks \cite{fn:ortho}. In this case, Eq.\ (\ref {waveeq})
immediately gives the dispersion relation
                \begin{eqnarray}
\omega^2 = {c^2 \over \epsilon\mu} k^2.   \label{20}
                    \end{eqnarray}
In non-dispersive media (i.e.\ when
$\epsilon$ and $\mu$ are independent of $\omega$ and $\vec k$), this
gives the speed of propagation as $c/\sqrt{\epsilon\mu}$.  Apart from
the transversality condition $\vec k \cdot \vec E = 0$, there is no
restriction on $\vec E$.  Therefore, we conclude that the two
transverse modes have the same physical characteristics and, in fact,
any linear combination of them is equally appropriate for discussing
the transverse wave properties.

In general, however, the two transverse modes have different
properties when $\zeta$ does not vanish. To proceed in this case, we
take the cross product of Eq.\ (\ref{waveeq}) with $\hat k$ and use
the transversality condition $\hat k \cdot \vec E = 0$ to obtain
                \begin{eqnarray}
\omega^2 \hat k \times \vec E = {c^2 \over \epsilon\mu} k^2 \hat k
\times \vec E + {c\zeta \over \epsilon} \omega k \vec E \,.
\label{kXwaveeq}
                \end{eqnarray}
Taking linear combinations, this pair of coupled equations for $\vec
E$ and $\hat k \times \vec E$ can be rewritten as
                \begin{eqnarray}
\left[\omega^2  - \left( {1 \over \epsilon\mu} \pm {i\zeta\omega \over
\epsilon kc} \right) c^2 k^2\right]
 \left( \vec E \pm i \hat k \times \vec E \right) = 0 \,.
                \end{eqnarray}
Therefore, solutions exist only if $\omega$ and $\vec E$ satisfy
                \begin{eqnarray}
\omega^2 &=&
\left( {1 \over \epsilon\mu} + \lambda  {i\zeta\omega \over \epsilon kc}
\right) c^2 k^2 \,, \label{dispersion} \\
\vec E &=& i \lambda \hat k \times \vec E \,,
\label{lrpol}
                    \end{eqnarray}
with $\lambda = \pm 1$.  Thus, the
two transverse modes no longer have the same dispersion relation.
Furthermore, an arbitrary linear combination of them does not, in
general, obey the wave equation (\ref{waveeq}). The combinations that
travel as waves are the ones that satisfy Eq.\ (\ref{lrpol}), which
are in fact the positive and negative circular polarization vectors
$\vec e \pm i\hat k\times\vec e$, where $\vec e$ is any vector
transverse to $\hat k$~\cite{fn:onlylr}.

We argued before, in connection with Eq.\ (\ref{Trules}), that
$\epsilon$ and $\mu$ are real whereas $\zeta$ is purely imaginary if
time-reversal invariance is satisfied. In this case, it follows from
Eq.\ (\ref{dispersion}) that that $\omega$ is real for real values of
$k$ if $\left|\zeta\right| < kc \left|\mu\right| /\omega$. The
amplitude of a monochromatic wave will thus remain unchanged during
the propagation, which means that there is no absorption in the
medium.  However, if $\zeta\neq 0$, the speeds of propagation of the
two circlularly polarized transverse modes are different.

In a more general case, time-reversal symmetry can be violated. Then,
for a given $\vec k$, $\omega$ is complex in general, so that the
exponential factors in Eq.\ (\ref {fourier}) give some damping (or
growth) with time. Physically, this corresponds to absorption by (or
emission from) the medium. Even in the case of $\zeta = 0$, absorption
of the two transverse modes can arise if $\epsilon$ or $\mu$ is
complex. But the absorption has to be equal for the two modes.  For
$\zeta \neq 0$, that is not the case.

For the sake of simplicity, we will stick to an unabsorbing medium.
This means that $\omega$ is real for all wave vectors $\vec k$. From
Eq.\ (\ref{dispersion}), this implies that $\epsilon$, $\mu$, are
real, $\zeta$ is purely imaginary and $\left|\zeta\right| < kc
\left|\mu\right| /\omega$.  As already remarked, in this simple case
the effect of a non-zero value of $\zeta$ is to produce a difference
in speed between the two circularly polarized degrees of freedom.

\section{Physical implications of $\zeta$}\label{implications}
            We now discuss the conclusions of the last section in a
more intuitive way on the basis of the discrete operations P, C and T.
Again, we concentrate on the transverse modes only, and draw the
polarization vectors $\hat e_1$, $\hat e_2$ along with the unit
wave-vector $\hat k$ in Fig.~\ref{actionP}a.

\figgap{
\begin{picture}(150,90)(10,5)
\thicklines
\put(20,50){\circle{5}}
\put(22.5,50){\vector(1,0){32.5}}
\put(20,52.5){\vector(0,1){32.5}}
\put(40,45){\makebox(0,0){{\large$\widehat{e}_2$}}}
\put(17,70){\makebox(0,0){{\large$\widehat{k}$}}}
\put(100,50){\circle*{5}}
\put(100,50){\vector(-1,0){35}}
\put(100,50){\vector(0,-1){35}}
\put(80,45){\makebox(0,0){{\large$\widehat{e}_2$}}}
\put(97,30){\makebox(0,0){{\large$\widehat{k}$}}}
\put(150,50){\circle{5}}
\put(147.5,50){\vector(-1,0){32.5}}
\put(150,52.5){\vector(0,1){32.5}}
\put(130,45){\makebox(0,0){{\large$\widehat{e}_2$}}}
\put(147,70){\makebox(0,0){{\large$\widehat{k}$}}}
\put(35,10){\makebox(0,0){{\Large\bf (a)}}}
\put(85,10){\makebox(0,0){{\Large\bf (b)}}}
\put(135,10){\makebox(0,0){{\Large\bf (c)}}}
\end{picture}}
{The effect of a Parity transformation on the wave vector $\vec k$ and
the transverse polarization vectors. The empty circles in diagrams (a)
and (c) indicate that the direction of $\hat e_1$ is towards the
reader, whereas the filled circle in diagram (b) indicates the
direction to be into the paper.} {actionP}
                                           Now imagine the effect of
parity on the vectors in Fig.~\ref{actionP}a. Surely $\hat k$ reverses
direction, since it is a polar vector denoting the direction of wave
propagation. To see the effect of parity on $\hat e_1$ and $\hat e_2$,
we first need a precise definition of them. This depends on the way we
choose the scalar and vector potentials to express the electric and
magnetic fields. For the sake of convenience, we take the ``radiation
gauge'' in which the scalar potential vanishes and
                \begin{eqnarray}
\vec E = i\omega \vec A, \quad \vec B = i\vec k \times \vec A\,.
\label{23}
                    \end{eqnarray}
In this gauge, we can express the vector potential as
                \begin{eqnarray}
\vec A = \sum_\lambda a_\lambda \hat e_{\vec k\lambda}  \,,
\label{24}
                    \end{eqnarray}
where the quantities $a_\lambda$
are numerical coefficients in classical electromagnetism. Now, under
parity, $\vec E$ reverses sign, so that from Eq.\ (\ref {23}), we see
that $\vec A$ must do the same. From Eq.\ (\ref {24}), we then obtain
that the vectors $\hat e_{\vec k\lambda}$ must also change sign. All
of these facts have been summarized in Fig.~\ref{actionP}b.

However, since we are dealing with isotropic systems, the absolute
directions of the vectors do not matter, only their relative
orientations do. Thus, rotating Fig.~\ref{actionP}b around the $\hat
e_2$-axis by $180^\circ$, we obtain Fig.~\ref{actionP}c.  Comparing
this with Fig.~\ref{actionP}a, we see that effectively, for the same
$\hat k$, the effect of the parity transformation has been to reverse
the sign of $\hat e_2$.

An alternative way to discuss the polarization of electromagnetic
waves is to use, instead of the linear polarization vectors $\hat e_1$
and $\hat e_2$, the right and left circular polarization vectors
defined by
                \begin{eqnarray}
\hat e_\pm \equiv {1 \over \sqrt 2} (\hat e_1 \pm i\hat e_2) \,.
                    \end{eqnarray}
Since $\hat e_2 = \hat k\times\hat
e_1$, these are of the form $\hat e_1\pm i\hat k\times\hat e_1$, as
mentioned after Eq.\ (\ref{lrpol}).  In this language, thus, the
effect of parity is easily understood.  Since effectively $\hat e_2
\longrightarrow -\hat e_2$, as shown earlier, we conclude that under
parity
                \begin{eqnarray}
\hat e_+ \longleftrightarrow \hat e_- \,.
                    \end{eqnarray}
Thus, if parity is a good symmetry,
there should be no difference in the physical properties of a
right-circular and a left-circular polarization state of an
electromagnetic wave. Such is the case, for example, in the vacuum,
since Eqs. (\ref{MaxEq1kw}) and (\ref{MaxEq2kw}) are invariant under
parity.

In matter, as we pointed out, this is not the case. Eq.\ (\ref
{jind-general}) and its corollary in Eq.\ (\ref {jext}) break parity
because of the $\zeta$-term.  Thus, in a medium, the right- and the
left- circularly polarized electromagnetic waves can have different
physical properties. For example, they travel with different speeds
through matter, have different indices of refraction and coefficients
of absorption.

It is easy to see that this difference manifests itself as optical
activity. Consider a plane polarized wave incident on a medium at
$x=0$, moving in the positive $x$-direction. Suppose the direction of
polarization of the incident wave is along $\varphi=0$, where
$\varphi$ denotes the azimuthal angle in the plane perpendicular to
the direction of propagation. As is well-known \cite{optics}, this
plane polarized wave can be treated as the superposition of a
right-circularly polarized wave with phase angle $\varphi_+(0)=\omega
t$ and a left-circular polarized wave of the same amplitude with phase
angle $\varphi_-(0)=-\omega t$. The zero in the paretheses here denote
the fact that these phases are given for the interface where $x=0$.

Now the wave propagates inside the medium, where $\zeta\neq 0$, and
consequently, the same value of $\omega$ corresponds to different
values of the wave number $k$ for the two circularly polarized
components, as we argued before. Let us call these values of $k$ by
$k_+$ and $k_-$.  Then, after traversing a distance $x$ within the
medium, the phase angle for the right- and left- circularly polarized
components become
                \begin{eqnarray}
\varphi_+(x) = \omega t - k_+ x, \quad \varphi_-(x) = -\omega t + k_- x
\,,
                    \end{eqnarray}
so that the resultant direction is given by
                \begin{eqnarray}
\varphi (x) = {1\over 2} \left[\varphi_+(x) + \varphi_-(x) \right]  = {1
\over 2} \left(  k_- - k_+ \right) x \,.
                    \end{eqnarray}
This shows that the direction of
polarization of the wave changes with $x$ within the medium, which is
precisely the phenomenon of optical activity.

A similar treatment can be applied to the phenomenon that occurs in
the optical-rotation experiments that have been performed to test the
parity-violating weak force between the electrons and nucleons, which
coexists with the electromagnetic force
\cite{parityviolation}.  In these
experiments, a beam of lineraly polarized light traverses a cell
containing atomic vapor of bismuth, and the direction of polarization
of the wave changes as the beam traverses the cell. From a macroscopic
point of view, the effect can be understood and described as discussed
above.

The last two equations, together with our earlier analysis leading to
Eq.\ (\ref{dispersion}), bring out clearly the fact that the
difference between $k_-$ and $k_+$ is introduced because $\zeta \neq
0$ \cite{fn:examples}.  In view of this, the quantity $\zeta$ may be
called the {\em activity constant} of a medium.

\section{Conclusions and summary}
              We have explained why, in order to describe
electromagnetic phenomena in an isotropic medium, one needs in general
three parameters, or electromagnetic constants. Two of them are
introduced and studied in detail in every textbook on Electromagnetic
Theory, and are called the {\em dielectric constant} and the {\em
magnetic permeability} of the medium in question.  The third one,
which we denote by the symbol $\zeta$ in this paper, is not as widely
discussed or not introduced at all in physics textbooks. We showed
that $\zeta$ has some features which are very different from the other
two. For example, a medium can have a non-zero value of $\zeta$ only
if there are asymmetries of parity, CP etc. These asymmetries may come
either from a violation of those symmetries in fundamental
interactions between particles, or from the constituents of the
medium. Since it is known that particle interactions violate parity
and CP, the value of $\zeta$ is in principle non-zero for all normal
media. Morever, if the constituents of the medium are
parity-asymmetric objects like helical molecules, the effect can be
large.

Physically, the presence of the $\zeta$ term introduces differences
between the characteristics of the right- and left-circularly
polarized electromagnetic waves. These differences manifest themselves
in various physical phenomena, e.g.\, optical activity and atomic
parity violation experiments. Our approach in the present paper
provides a general framework to study and interpret these phenomena
from a unified point of view.

\appendix
\part*{Appendix}
\section{Time reversal of the Maxwell equations in Classical and Quantum
mechanics} As we have discussed in the text, a non-zero value of
$\zeta$ has deep implications regarding the discrete space-time
symmetries of the system.  The transformation rule that we have used
for Time-Reversal is apropriate in the context of Classical Physics
which, for clarity, we have adopted in the text. In the realm of
Quantum Mechanics that definition of Time-Reversal cannot be used.
Therefore, it is important to discuss the appropriate transformation
rule in Quantum Mechanics, since in principle $\zeta$ has to be
calculated in that context.  In this appendix, we discuss the
different definitions of the time-reversal transformation and why they
arise.

Let us use as starting point Eq.\ (\ref{jext}) which gives the
external current density in terms of the electric and magnetic fields.
For the sake of convenience, we define the following quantities:
                \begin{eqnarray}
\label{eq2}
L_E^{ab} (\vec k, \omega) & = & i\omega\epsilon \delta^{ab}\nonumber\\
L_B^{ab} (\vec k, \omega) & = & {c\over\mu} i \varepsilon^{acb} k^c +
i\omega\zeta \delta^{ab}\,,
                \end{eqnarray}
where the superscripts $a$,
$b$ denote Cartesian components, $\delta^{ab}$ denotes the Kronecker
delta symbol and $\varepsilon^{acb}$ is the completely antisymmetric
tensor with $\varepsilon^{123}=+1$. With these, we can rewrite Eq.\
(\ref{jext}) as
                \begin{eqnarray}
\label{eq3}
\vec j^a \ext = \sum_b L_E^{ab} \vec E^b  + L_B^{ab} \vec B^b \,.
                    \end{eqnarray}
Since $\vec j\ext$ and $\vec B$
change sign under time reversal whereas $\vec E$ does not, we can say
that the Time-Reversal invariance of the Maxwell equations is
equivalent to the statement that $L_E$ and $L_B$ have the following
properties:
                \begin{eqnarray}
\label{eq4}
L_E(\vec k, -\omega ) & = & -L_E(\vec k, \omega)\nonumber\\ L_B(\vec
k, -\omega) & = & L_B(\vec k, \omega) \,.
                \end{eqnarray}
These in turn imply the relations on $\epsilon$, $\mu$ and $\zeta$
that have been obtained in the text.  This definition of the
Time-reversal operation can be called the classical definition.  If
Eq.\ (\ref{eq4}) holds, then the Maxwell equations are invariant under
the substitutions
                \begin{eqnarray}
\label{eq5}
\vec E (\vec k, \omega) \stackrel{T}{\longrightarrow} &
\vec E' (\vec k ',\omega')  & \equiv \vec E (\vec k, \omega) \,,
\nonumber\\
\vec B (\vec k, \omega) \stackrel{T}{\longrightarrow} & \vec B'
(\vec k ',\omega') & \equiv -\vec B (\vec k, \omega) \,,
\nonumber\\
\vec j (\vec k, \omega) \stackrel{T}{\longrightarrow} &
\vec j'(\vec k ',\omega') & \equiv -\vec j (\vec k, \omega) \,,
\label{TclassEBj}
                    \end{eqnarray}
where
                \begin{eqnarray}
(\vec k, \omega) \stackrel{T}{\longrightarrow} & (\vec k ',\omega') &
\equiv (\vec k, -\omega) \,.
\label{eqTclass}
                    \end{eqnarray}
Eq. (\ref{TclassEBj}) is, of
course, equivalent to the identification of the value of $\eta_T$ for
the various quantities given in Table~\ref{t:CPT}.

In Quantum Mechanics, however, these transformation rules cannot be
used. The reason is that, while they leave the Maxwell equations
invariant, these rules do not leave invariant the quantum mechanical
equations of motion for the charges, namely the Schrodinger equation.
A simple way to understand this is to recall that, in Quantum
Mechanics, one identifies $\omega={\cal E}/\hbar$ and $\vec k=\vec
p/\hbar$, where ${\cal E}$ and $\vec p$ denote the energy and momentum
of the particle and $\hbar$ is Planck's constant divided by $2\pi$.
Now, from the non-relativistic definitions of momentum and kinetic
energy of a point particle [viz., $\vec p=m {d\vec r\over dt}$ and
${\cal E}\sub{kin} = {1\over 2} m
\left( {d\vec r\over dt} \right)^2$], one expects that $\vec p$
changes sign under time-reversal whereas ${\cal E}$ does not. Thus,
instead of Eq. (\ref{eqTclass}), the transformation is defined by
                \begin{eqnarray}
(\vec k, \omega) \stackrel{T}{\longrightarrow} &
(\vec k',\omega') & \equiv (-\vec k,\omega) \,.
\label{eqTqm}
                    \end{eqnarray}
This is how the transformation must
be defined in order to leave the Schrodinger equation for a free
particle invariant.  The interesting point is that this transformation
is a symmetry operation of the Maxwell equations also, provided that
$L_E$ and $L_B$ satisfy, instead of Eq.\ (\ref{eq4}), the following:
                \begin{eqnarray}
\label{eq6}
L_E^\ast(-\vec k,\omega) & = & -L_E(\vec k, \omega)\,,\nonumber\\
L_B^\ast(-\vec k,\omega) & = & L_B(\vec k, \omega) \,.
                \end{eqnarray}
If these relations hold, then the Maxwell equation is invariant under
the substitutions:
                \begin{eqnarray}
\vec E (\vec k, \omega) \stackrel{T}{\longrightarrow} & \vec E'
(\vec k',\omega') & \equiv \vec E^\ast(\vec k, \omega)\nonumber\\
\vec B (\vec k, \omega) \stackrel{T}{\longrightarrow} &
\vec B' (\vec k',\omega') & \equiv -\vec B^\ast(\vec k,
\omega)\nonumber\\
\vec j\,(\vec k, \omega)
\stackrel{T}{\longrightarrow}  & \vec j\:'(\vec k',\omega') &
\equiv -\vec j^\ast(\vec k, \omega)\nonumber\\
i \stackrel{T}{\longrightarrow} & i' & \equiv i^\ast = -i \,.
\label{eq7}
                    \end{eqnarray}
We will refer to these
transformation rules as the Quantum Mechanical ones.  Since the
transformation rules are different from the ones discused before, the
consequences of time reversal symmetry on $\epsilon$, $\mu$ and
$\zeta$ change as well.

The last comment is meant to emphasize that Eqs.\ (\ref{eq5}) and
(\ref{eq7}) are two different transformation rules, and the symmetries
that they imply are also different. However, they are equivalent if,
but only if, $L_E$ and $L_B$ do not have an absorptive part. In that
case, those quantities depend on $i$ only through the combinations
$i\vec k$ and $i\omega$, and therefore taking the complex conjugate of
them is the same as changing the sign of $\omega$ and $\vec k$ in
their argument. It is then immediately seen that Eqs.\ (\ref{eq4}) and
(\ref{eq6}) are really the same. However, if $L_E$ and $L_B$ have an
absorptive part, then the two transformation laws are different.

The classical transformation is a symmetry operation only if Eq.\
(\ref{eq4}) holds. That equation implies the relations which have been
derived in Eq.\ (\ref{Trules}), which in turn imply that $\epsilon$,
$\mu$ are real while $\zeta$ is purely imaginary; i.e., neither have
an absorptive part. The classical transformation law is not a symmetry
operation if any of those quantities has an absorptive part.  However,
the quantum mechanical transformation law is a symmetry operation even
when they have an absorptive part; what is needed is that Eq.\
(\ref{eq6}) be satisfied, and that is always true because $\epsilon$,
$\mu$ and $\zeta$ satisfy the relation (\ref{hermiticity}). As already
discused, in classical physics this relation is a consequence of the
fact that the fields and the current are real quantities, while in
Quantum Mechanics it is just the requirement of Hermiticity of the
current and the fields.

We summarize the situation as follows.  In a non-absorbing medium, the
classical and quantum definitions of the Time-Reversal operation
coincide.  However, the classical Time-Reversal symmetry is broken if
$\epsilon$, $\mu$ or $\zeta$ have an absorptive part.  Therefore, in
the realm of Classical Physics, it is impossible to get an absorptive
part of those quantities except if the theory breaks the (classical)
Time-Reversal symmetry. The Schrodinger equation breaks that symmetry;
quantum effects break that transformtion. But the equations of Quantum
Mechanics (and of course Maxwell equations) are invariant under a more
general transformation, which continues to hold even when $\epsilon$,
$\mu$ and $\zeta$ have an absorptive part.  The existence of the
$\zeta$ term implies that P is broken but not T (with the quantum
mechanical definition of T) nor C. Thus, $\zeta$ breaks P, CP and CPT.


\begin{thebibliography}{[00]}

\bibitem{text}
See any text in electrodynamics, e.g., E.~M. Purcell, {\em Electricity
and Magnetism}, (Berkeley Physics Course, vol. 2; McGraw-Hill, New
York 1989), Chaps. 10--11; D. Halliday and R.  Resnick, {\em
Fundamentals of Physics} (John Wiley \& Sons, New York 1986) Chap. 27.

\bibitem{OpAct} See, for example,
E. U. Condon, {\em Theories of optical rotatory power}, Rev. Mod. Phys.
9, 432--457 (1937); P. Drude, {\em The theory of optics}, (Dover, 1959);
E.~J. Post, {\em Fundamental Structure of Electromagnetics},
(North-Holland, 1962); E. Charney, {\em The molecular basis of optical
activity} (Wiley, New York 1979);
A. Lakhtakia, V.~K. Varadan and V.~V. Varadan, {\em Time-Harmonic
Electromagnetic fields in Chiral Media}, (Springer, 1989); J.~A. Kong,
{\em Electromagnetic Wave Theory}, (John Wiley \& Sons, 1986).

\bibitem{physicsbooks}
There are exceptions, e.g., the book by L. D. Landau and E. M.
Lifshitz, {\em Electrodynamics of Continuos Media, Course of
Theoretical Physics, Volume 8}, (Pergamon Press, 1960) pp 337.

\bibitem{NP1} J.~F. Nieves and P.~B. Pal, {\em P- and CP- odd terms in
the photon self-energy within a medium}, Phys. Rev. D 39, 652--659
(1989).

\bibitem{NP2} J.~F. Nieves and P.~B. Pal, {\em Propagation of gauge
fields within a medium}, Phys. Rev. D 40, 1350--1353 (1989).

\bibitem{fn:dim}
More commonly, one expresses the dimension of charge can be expressed
in terms of
the fundamental dimensions of length, time and mass through the
Coulomb force law. However, since the mass dimension is not directly
necessary for our purpose, we use length, time and charge as the
fundamental units in this paper.

\bibitem{fn:fourier}
In general, what we need to do is to look for solutions which are
expressed as a superposition of plane waves, with $\vec k$ and
$\omega$ not being necessarily real. Such waves are called {\em
inhomogeneous plane waves}, and they form a general basis for the
treatment of boundary-value problems for waves.  Those waves in
general exhibit exponential growth or decay in time, and also in space
along some directions, some simple examples being the phenomena of
total internal reflection and refraction in conducting media.  For our
present concerns, we are considering a spatially infinite medium, in
which case $\vec k$ must be taken to be real, but $\omega$ must in
general be allowed to be complex.  Therefore, in Eq.~(\ref{fourier}),
$\vec k$ is real but $\omega$ can be complex, and the integral over
$\omega$ must be understood accordingly.  For further discussion about
these points see, e.g., J. D. Jackson, {\em Classical
Electrodynamics}, (John Wiley \& Sons, New York 1975) Chap. 7.

\bibitem{fn:tensor}
In the most general case $\epsilon$ and $\mu^{-1}$ are rank-2 tensors,
and the terms denoted by 1 in Eq. (\ref{P&M}) stand for the unit tensor.

\bibitem{Monzon}
J. C. Monzon, {\em Radiation and scattering in homogenoeous general
biisotropic regions}, IEEE Trans. Ant. Prop. 38, 227--235 (1990).

\bibitem{Krowne}
C. M. Krowne, {\em Electromagnetic theorems for complex anisotropic
media}, IEEE Trans. Antennas Propag. 32, 1224--1230 (1984).

\bibitem{Kong} J. A. Kong, {\em Theorems of bianisotripic media},
Proc. IEEE 60, 1036--1046 (1972).

\bibitem{Tellegen}
B. D. H. Tellegen, {\em The gyrator, a new network element}, Phillips
Res. Rept. 3, 81--101 (1948).

\bibitem{Satten} See, e.g., R. A. Satten, {\em Time Reversal Symmetry
and Electromagnetic Polarization Fields}, J. Chem. Phys. 28, 742--743
(1958).

\bibitem{fn:T} We are discussing the Time-Reversal operation
here in classical terms, and we will keep that spirit in the rest of
the paper. In the appendix we discuss the definition that must be
adopted in the realm of Quantum Mechanics and also show that in a
non-absorbing medium the two definitions coincide.

\bibitem{fn:proton}
Protons might be present in order to attain a neutral medium.
However, their presence does not affect our conclusion.

\bibitem{fn:ortho}
In Eq.~(\ref{waveeq}) it cannot be assumed that $\hat k\times\vec E$
and $\vec E$ are linearly independent, and therefore we cannot
conclude that the coefficient of the last term must be zero.  For a
complex vector $\vec E$, $\hat k\times\vec E$ and $\vec E$ can be
linearly dependent.

\bibitem{fn:onlylr}
In fact, it is easy to prove (see, e.g., Condon, Ref.\
\cite{OpAct}) that the left- and right- circular polarized waves are
the only ones which can travel through the medium with unchanged state
of polarization. For this, write Eqs.\ (\ref{waveeq}) and
(\ref{kXwaveeq}) in a matrix form:
                \begin{eqnarray*}
\left( \begin{array}{cc}
\omega^2 - {c^2 \over \epsilon\mu} k^2 &
{c\zeta \over \epsilon} \omega k \\ - {c\zeta \over \epsilon} \omega k
& \omega^2 - {c^2 \over \epsilon\mu} k^2 \end{array} \right)
\left( \begin{array}{c}
\vec E \\   \hat k \times \vec E
\end{array} \right) = 0 \,.
        \end{eqnarray*}
This is an eigenvalue problem whose
characteristic equation yields the dispersion relations of
Eq.\ (\ref{dispersion}), with the eigenvectors determined as the
solutions of Eq.\ (\ref{lrpol}).

\bibitem{optics}
Various textbooks on Optics discuss this. See, e.g., B. Rossi, {\em
Optics}, (Addison-Wesley, Reading 1957) Chap. 6.

\bibitem{parityviolation}
See, e.g., E. D. Commins and P. H. Bucksbaum,
{\em Weak Interactions of leptons and quarks}, (Cambridge
University Press 1983) pp 349.

\bibitem{fn:examples}
Explicit solutions of Eq.~(\ref{dispersion}), which clearly illustrate
how the dispersion relations for right- and left-circularly polarized
waves differ, can be readily obtained for the cases in which
$\epsilon$ and $\mu$ are constants and $\zeta$ is proportional to
either $\omega$, $\omega^{-1}$, $\omega^{3}$ or $\omega^{-3}$.  On the
other hand, perturbative solutions can be obtained in the case
that $\zeta$ is small.

\end{thebibliography}
\end{document}